# Masses of Nix and Hydra

proposed running head: Masses of Nix and Hydra

DAVID J. THOLEN
*Institute for Astronomy, University of Hawaii, 2680 Woodlawn Drive, Honolulu, HI 96822*
tholen@ifa.hawaii.edu

MARC W. BUIE AND WILLIAM M. GRUNDY
*Lowell Observatory, 1400 W. Mars Hill Road, Flagstaff, AZ 86001*

GARRETT T. ELLIOTT
*The Ohio State University, Columbus, OH 43210*

14 text pages
6 figures
6 tables

ABSTRACT

A four-body orbit solution for the Pluto system yields GM values of 870.3 ± 3.7, 101.4 ± 2.8, 0.039 ± 0.034, and 0.021 ± 0.042 km$^3$ sec$^{-2}$ for Pluto, Charon, Nix, and Hydra, respectively. Assuming a Charon-like density of 1.63 gm cm$^{-3}$, the implied diameters for Nix and Hydra are 88 and 72 km, leading to visual geometric albedos of 0.08 and 0.18, respectively, though with considerable uncertainty. The eccentricity of Charon's orbit has a significant nonzero value; however, the 0.030 ± 0.009 deg yr$^{-1}$ rate at which the line of apsides precesses is insufficient to explain the difference in the longitude of periapsis seen in the orbits fitted to the 1992-1993 and 2002-2003 data sets. The mean orbital periods for Hydra, Nix, and Charon are in the ratios of 6.064 ± 0.006 : 3.991 ± 0.007 : 1, but we have not identified any resonant arguments that would indicate the existence of a mean motion resonance between any pairs of satellites.

1. INTRODUCTION

	The discovery of two new satellites of Pluto (Weaver *et al.* 2005, 2006) made this system even more interesting than it already was.  Two-body orbit solutions for Nix and Hydra were computed by Buie *et al.* (2006, hereafter referred to as B06) using prediscovery observations of the new satellites.  Their use of the Pluto-Charon barycentric reference frame implicitly took into account the indirect perturbations by Charon (on Pluto), but not the direct perturbations (on Nix and Hydra), which manifested themselves as inconsistent values for the sum of Pluto's and Charon's masses (Lee & Peale 2006).  Accounting for the indirect perturbations resulted in a rather accurate determination of the Charon / Pluto mass ratio, a quantity that several authors have struggled to obtain from measurements of the barycentric wobble (Null *et al.* 1993; Young *et al.* 1994; Null & Owen 1996; Tholen & Buie 1997; Foust *et al.* 1997; Olkin *et al.* 2003).
	Although the two-body orbit solutions are good enough to satisfy the bulk of the 2002-2003 data, we were concerned that the direct perturbations might be too strong to permit a sufficiently accurate extrapolation forward to the 2015 New Horizons spacecraft encounter with Pluto, backward to the 1985-1990 Pluto-Charon mutual events (we wondered whether the new satellites might have been detected in the photometric data), or to predict future stellar occultations.  Also, with adequate data, a formal four-body orbit solution should yield the mass for each member of the system.  With these goals in mind, we performed such an orbit solution using the same set of observations.

2. DATA

	Our four-body orbit solution was performed using astrometric data for Charon, Nix, and Hydra from several sources.  For Nix and Hydra, we used the 12 positions from B06, the two discovery positions from Weaver *et al.* (2006), and the two follow-up positions from Stern *et al.* (2007), all obtained with the Hubble Space Telescope.  Additional ground-based observations of both new satellites from the Magellan telescope (Fuentes & Holman 2006) and for Hydra from the VLT (Sicardy *et al.* 2006a) were checked for consistency, but were not used in the orbit solution for reasons described below.  For Charon, we used the 60 positions from Tholen & Buie (1997, hereafter referred to as TB97), the 384 positions extracted from the individual exposures described in (but not tabulated by) B06, and the two positions accompanying the discovery observations of Nix and Hydra mentioned above.  To constrain the orbital period of Charon more tightly, we included average positions from 1985 April 27-29 UT, the three nights of speckle interferometric data showing the least scatter (Beletic *et al.* 1989).
	Before data sets from different sources can be readily combined, it is important to determine whether there are any significant systematic differences in their scale and orientation calibrations.  The TB97 and B06 observations of Charon are of particular concern because the fitted semimajor axes of $19636 \pm 8$ and $19571 \pm 4$ km differ by over 7$\sigma$.  We have considered three possible explanations for the discrepancy:  (a) one or both of the image scale determinations is in error, (b) the TB97 semimajor axis suffers from a systematic error due to the offset between the center of mass and the center of light, or (c) the osculating semimajor axis varies by at least 40 km in response to perturbations by Nix and Hydra. Without knowing the correct explanation prior to beginning the orbit solution process, we simply chose to inflate the astrometric uncertainties on both data sets so that the reduced $\chi^2$ statistic was close to unity when a two-body orbit solution for Charon was performed on the combined data. The assigned uncertainty for all 384 B06 observations was increased from 0.0019 to 0.0022 arcsec. The 60 TB97 observations had two uncertainty levels associated with them, namely 0.0060 arcsec for the six observations made closest to minimum separation, and 0.0024 arcsec for the remaining 54 observations; these were increased to 0.0070 and 0.0028 arcsec, respectively. It should be noted that even though these adjusted uncertainties produce the desired reduced $\chi^2$ statistic, the sheer number of B06 observations dominates the solution for the semimajor axis of the orbit when using the combined data set. The calibration for the discovery observations was not specified by Weaver *et al.* (2006); however, the stated uncertainty of 0.035 arcsec is large enough, and the number of Charon observations is small enough, that a systematic error in the calibration would likely be undetectable. The semimajor axis for Charon's orbit computed by Beletic *et al.* (1989) has a sufficiently large uncertainty to be compatible with both the TB97 and B06 calibrations, so we adopted astrometric uncertainties equal to the standard deviation of the positions for Charon on each of the three nights utilized.

B06 assigned uniform astrometric uncertainties of 0.015 arcsec to the observations of Nix and 0.009 arcsec to the observations of Hydra. However, they were forced to reject one observation of Nix and two observations of Hydra due to excessive residuals. Our initial four-body orbit solution showed that it was possible to satisfy two of those three rejected observations, though they were sufficiently noisy to inflate the reduced $\chi^2$ statistic. We compensated for that inflation by increasing the astrometric uncertainties for Nix and Hydra to 0.017 and 0.010 arcsec, respectively. The uncertainties for the Weaver *et al.* (2006) positions for Nix and Hydra were set at 0.035 arcsec, and the Stern *et al.* (2007) astrometry has stated uncertainties of 0.010 arcsec for the data taken on 2006 February 15 UT and 0.015 arcsec for the data taken on 2006 March 2 UT.

## 3. ORBIT SOLUTION

To perform the orbit solution, we used the 15th order Gauss-Radau numerical integrator developed by Everhart (1985). Because his code was designed to operate in the heliocentric reference frame and accounts for the indirect terms, we chose to work in the Plutocentric reference frame, with the central mass being one of the solution parameters. The other 21 solution parameters involved the mass, position vector, and velocity vector for each of the three satellites. Because all three satellites are deep in Pluto's gravitational well, we ignored perturbations by the Sun, as did Lee & Peale (2006). We evaluated the speed of two different function minimization routines, namely the downhill simplex method (see page 289 of Press *et al.* 1986) and the gradient search method (see page 215 of Bevington 1969). Both routines start by evaluating the slope of the $\chi^2$ hypersurface in each dimension. The downhill simplex method then manipulates the vertex of the simplex having the largest $\chi^2$ value, while the gradient search method takes repeated steps along the same gradient until the $\chi^2$ statistic starts to increase, performing a parabolic interpolation of the final three steps. Neither routine demonstrated any significant superiority over the other overall, though there were certain circumstances in which each did outperform the other. We kept iterating until such time that after a restart, the output values agreed with the input values to $10^{-3}$ km in position, $10^{-5}$ km day$^{-1}$ in velocity, and $10^{-7}$ km$^3$ sec$^{-2}$ in GM.

All 22 parameters were fitted simultaneously, with the initial position and velocity vectors taken from the two-body orbit solutions for each object, and those for Nix and Hydra converted to the Plutocentric reference frame. The initial masses for Pluto and Charon were based on the system mass derived from the Plutocentric two-body orbit of Charon and the mass ratio of 0.1165 from B06. The initial masses for Nix and Hydra were constrained by coupling the available photometry with a range of plausible geometric albedos from 0.04 (like some comets) to 0.64 (similar to Pluto), which produce diameters spanning a factor of 4, thus volumes spanning a factor of 64, and then a range of plausible densities from 1.0 (water ice) to 2.0 (like Pluto) gm cm$^{-3}$, which in turn produces a range of masses spanning a factor of about 130. The resulting masses range from about $2.0 \times 10^{16}$ to $2.6 \times 10^{18}$ kg, with the corresponding GM values ranging from about 0.001 to 0.17 km$^3$ sec$^{-2}$. Although densities more extreme than these have been observed in the Solar System, the four-body solution in no way forced the masses to fall within any particular range. Our goal was simply to find a reasonable set of starting values. We started the solution at both ends of the range of masses (high-high and low-low) and verified that they both converged to the same final values; we did not test the high-low and low-high mass permutations for Nix and Hydra.

The 22-dimensional $\chi^2$ hypersurface for the solution involving this data set is extremely complex. While performing the computations, the solution settled into many different local minima. We always restarted the solution by taking steps of various sizes (increasing by factors of two as many as eight times) in each of the parameters involving Nix and Hydra, sometimes finding a deeper minimum in the vicinity of that new starting point. That process was repeated until the steps no longer caused the solution to migrate toward a deeper minimum. So although we have performed an extensive examination of parameter space, it is not possible to guarantee that we have found the absolute minimum without computing an impractical number of orbit solutions, given the 22-dimensional nature of the problem.

The four-body orbit solution (with very tight convergence tolerances as noted above), the subsequent extensive exploration of parameter space to determine whether the adopted solution was in a local minimum rather than an absolute minimum, and the computation of the 1σ uncertainties in each parameter, was extremely computationally intensive. The process was well underway before the ground-based observations of Fuentes & Holman (2006) and Sicardy *et al.* (2006a) became available, which is the primary reason for not including them in the solution. Furthermore, those observations are sufficiently small in number that it is not possible to test whether the image scale and orientation calibrations are consistent with the HST observations. Lastly, they extend the observational arc for Nix and Hydra by only 117 days, less than ten percent of the total arc. We therefore chose not to restart the orbit solution and the exploration of parameter space when these observations became available. We did, however, check them for consistency with our orbit solution. The residuals are discussed below.

We also repeated the three two-body solutions in B06 using the exact same set of observations and our newly adopted astrometric uncertainties. The $\chi^2$ statistic of the four-body solution is smaller than the sum of the statistics from the three two-body solutions at the 2.3σ level.

The state vectors resulting from our four-body orbit solution are shown in Table 1 along with the 1σ uncertainties on each of the values, which appear directly above. That the uncertainties for Nix and Hydra are generally one to two orders of magnitude larger than those for Charon reflects the paucity of data for the new satellites. We adopted the same epoch as B06, which falls near the middle of their extensive 2002-2003 data set. All computations were done in a Plutocentric reference frame aligned with Earth's mean equator and equinox of J2000.

Table 2 shows the GM values that resulted from our four-body orbit solution. If we adopt a value of $6.67428 \times 10^{-11}$ m$^3$ kg$^{-1}$ sec$^{-2}$ for the Newtonian gravitational constant (CODATA 2006), the corresponding masses are as shown in the table. The diameters of Pluto and Charon were determined from fits to the photometry of mutual events (Tholen & Buie 1990; Young & Binzel 1994; Reinsch *et al.* 1994), although the diameter of Charon was more recently measured to higher accuracy via the stellar occultation method (Young *et al.* 2005; Gulbis *et al.* 2006; Sicardy *et al.* 2006b; a synthesis of all three data sets was performed by Person *et al.* 2006). The thin atmosphere of Pluto prevents the stellar occultation method from seeing the solid surface of Pluto, so even though occultations have now been observed in 1988, 2002, 2006, and 2007, only an upper limit on the diameter of Pluto has been derived from that technique, therefore we have chosen to adopt the Tholen & Buie (1990) mutual event result here. However, because the mutual event size determinations scale according to the adopted semimajor axis of Charon's orbit, we are using 2294 km for Pluto's diameter to correspond to the new 19570 km value for the semimajor axis. The uncertainty in the diameter of Pluto is the largest contributor to the uncertainty in its density. It should be emphasized that the albedo for Pluto shown in the table is a globally averaged value; Pluto is known to exhibit substantial surface contrast.

To estimate the diameters of Nix and Hydra, we have assumed a Charon-like density of 1.63 gm cm$^{-3}$ (square brackets are used in the table to indicate assumed quantities). The best-fit masses thus correspond to diameters of 88 and 72 km, respectively. When combined with the apparent magnitudes from Stern *et al.* (2007) and a Charon-like phase coefficient (Buie *et al.* 1997), we computed absolute magnitudes of 8.6 and 8.2, respectively, which imply visual geometric albedos of 0.08 for Nix and 0.18 for Hydra, though with considerable uncertainty, which will be discussed in more detail below.

For ease of interpretation and to facilitate comparison with the two-body results of B06, we show the Keplerian orbital elements for the three satellites in Table 3. In all cases, the mass used in the computation of the semimajor axis, mean longitude, and orbital period included the mass of the body in question and the masses of all objects interior to the orbit of that body. The semimajor axis and eccentricity are Plutocentric for Charon and barycentric for Nix and Hydra. Because the osculating values for some of the quantities vary in ways that would not produce a normal distribution, we have chosen not to attach formal statistical uncertainties to these values.

Although the orbit of Charon is well described by the Keplerian elements in Table 3, the orbits of Nix and Hydra are much more strongly perturbed, so in Table 4, we show the mean orbital elements for the three satellites, averaged over a fifty year interval, with the Keplerian elements computed at quarter-day intervals. The poles of the orbits for Nix and Hydra precess around the pole of the system's invariable plane with periods of approximately 5 and 15 years, respectively, therefore their mean inclinations and ascending nodes are the same as Charon's. The arguments of periapsis for Nix and Hydra are meaningless because the value circulates so rapidly for each satellite. The uncertainties shown represent the standard deviations of 25 sets of mean orbital elements generated from a fifty-year integration for each of the 1σ parameter sets (see next section).

The orbit solution residuals for Nix and Hydra are shown in Table 5. The 2002 June 18 UT position of Hydra was not used for the orbit solution due to an excessive residual in right ascension. Unlike B06, we did retain the 2002 October 03 UT position of Nix, despite a large residual in declination, as well as the 2003 June 08 UT position of Hydra, whose residuals are quite acceptable. We have included the 2006 April and June ground-based observations in the table to show their consistency with our orbit solution, but we reiterate that they were not included in the least-squares fit. The estimated astrometric uncertainties for the April and June observations are 0.02 arcsec (Sicardy *et al.* 2006a) and 0.1 arcsec (Fuentes & Holman 2006), respectively, so the residuals demonstrate that those observations are consistent with the four-body solution at the 1.7$\sigma$ level or better. All positions excluded from our four-body orbit solution have had their residuals enclosed in parentheses in Table 5.

Table 5

## 4. UNCERTAINTIES

The uncertainties in the various solution parameters are usually an indication of the shape of the minimum on a multi-dimensional $\chi^2$ hypersurface. Unfortunately, the currently available data are insufficient to produce a single well-defined minimum. Indeed, the topography of the hypersurface is so complex that there is no reason to believe that the 3$\sigma$ uncertainty will be three times the 1$\sigma$ uncertainty, nor is the minimum necessarily symmetric in each dimension. With that said, we have computed the 1$\sigma$ uncertainties in each parameter, such that when the uncertainty is added to the parameter in question, and the other 21 parameters are allowed to adjust themselves to achieve a minimum in the $\chi^2$ statistic, the increase in that statistic over the best-fit value is approximately unity. The results are shown above each vector component in Table 1 and alongside the masses in Table 2. It should be noted that if one attempts to reproduce these values by adding the entire 1$\sigma$ value and allowing some $\chi^2$ minimization procedure to converge, there is a high probability that the solution will settle in a local minimum whose $\chi^2$ value is much greater than the best-fit value by considerably more than unity, potentially giving the illusion that the uncertainties have been overstated. Our procedure for finding the uncertainties involved incrementing the value of the parameter in question by a small amount, allowing the minimization procedure to follow valleys in the hypersurface as we gradually increased the offset in the parameter from the best-fit value. We used 25 different CPUs simultaneously to determine the 1$\sigma$ uncertainties in the 22 solution parameters and three other quantities of interest (described below). As is the case for the best-fit solution, we cannot guarantee that each solution for the uncertainty in the parameter being tested is in the absolute minimum, thus the stated uncertainty could be underestimated. Also, as of this writing, we have tested only positive increments to the values of most of the solution parameters. The 1$\sigma$ uncertainties for negative increments could be larger or smaller, if the minimum is asymmetric. We are confident, however, that we have not overestimated any uncertainty. Clearly, more data are needed to simplify the shape of the $\chi^2$ hypersurface and reduce the sizes of the uncertainties.

In addition to the 22 solution parameters, we were also interested in knowing the uncertainties in the total mass of the system as well as the Charon / Pluto mass ratio. To estimate these uncertainties, it was necessary to hold two solution parameters at fixed values while allowing the other 20 parameters to optimize themselves. In such cases, the appropriate increase in the value of the $\chi^2$ statistic is not unity, but rather 2.30 (see page 536 of Press *et al.* 1986). To estimate the uncertainty in the total mass of the system, which is dominated by the masses of Pluto and Charon, we incremented the masses of both Pluto and Charon such that the mass ratio was held constant until the $\chi^2$ statistic reached the target value (the best-fit value plus 2.30). The uncertainty in the total mass shown in Table 2 is technically the uncertainty in the sum of just Pluto's and Charon's masses, but because the uncertainty is much larger than the combined solution masses for Nix and Hydra, it is also effectively the uncertainty in the total mass of the system. To estimate the uncertainty in the Charon / Pluto mass ratio, we held the sum of their masses constant, thus requiring the increments on the masses of Pluto and Charon to be of equal magnitude but of opposite sign. Unlike the other 23 uncertainties, we tested both positive and negative increments on the mass of Charon (with increments of opposite sign on the mass of Pluto) to determine upper and lower limits on the mass ratio, giving us two more solution sets for the total of 25 mentioned above.

As one might expect, several parameters are strongly correlated with one or more of the other parameters. Most notably, the orbits of Nix and Hydra around the system barycenter tightly constrain the sum of the Pluto and Charon masses, but are not nearly as effective at constraining their individual masses, therefore the mass of Pluto is strongly anticorrelated with the mass of Charon. This situation is reflected in the tabulated 1$\sigma$ uncertainties, in which the total mass of Pluto and Charon (shown as the uncertainty in the total mass of the system) is known better than either of their individual masses.

The uncertainties reflect the random error in the data but not any potential systematic error. For example, the uncertainties in the position vector components for Charon are consistent with the 4 km uncertainty in the semimajor axis of Charon's orbit computed by B06. However, the dimensions of the Pluto system depend rather critically on the determination of the image scale of the ACS HRC. As we noted earlier, the semimajor axes determined for Charon by TB97 and B06 differ by 7$\sigma$, which directly corresponds to a similar discrepancy in the total GM of the system (981.5 ± 1.1 km$^3$ sec$^{-2}$ determined by TB97 and 971.78 ± 0.20 km$^3$ sec$^{-2}$ derived from this work). Our evaluation of the various possibilities for the discrepancy is in the Discussion section below.

5. MUTUAL EVENTS

Because the orbits for Charon, Nix, and Hydra are effectively coplanar, the mutual event seasons for all three satellites essentially coincided. The greater distances and smaller sizes of Nix and Hydra reduced the lengths of their seasons to about one-third the duration of Charon's. One of our goals is to determine whether any mutual events involving Nix or Hydra might have been serendipitously observed during one of the events involving Charon. The photometric detectability of either a Nix or Hydra event would be possible, though marginal, in the best data given the size estimates derived here. In particular, there are some subtle features in the mutual event lightcurve of 1988 April 18 UT (see Fig. 2 of Tholen & Buie 1988) that were originally attributed to possible post-eclipse brightening, but the discovery of two new satellites offers a possible alternative explanation.

Extrapolating the current orbits (and their uncertainties) for the new satellites back to the late 1980s indicates a timing uncertainty of almost an hour in the occurrence of a mutual event. Better orbits would facilitate searching the photometry archive for possible Nix or Hydra events.

Because we expect to have improved orbits in the near future, we decided to postpone this investigation until then.

## 6. DISCUSSION

*Satellite masses.* The origin of Nix and Hydra is an important question that may have far-reaching implications for the outer Solar System. Because all three satellites have similar colors (Stern *et al.* 2007), there is now an expectation that all are compositionally similar. If the models of Ward and Canup (2006) are correct, these objects coalesced from the products of the proto-Pluto and proto-Charon binary-forming impact. Their geometric albedos and bulk densities are probably the most diagnostic quantities. These values for Charon are well in hand but such determinations for Nix and Hydra will have to wait for stellar occultation observations, or, at worst, the New Horizons flyby in 2015. Nonetheless, our newly determined masses for Nix and Hydra now let us place some interesting constraints on this satellite system.

If we assume a Charon-like density of 1.63 gm cm$^{-3}$ for Nix and Hydra, we derive diameters of 88 and 72 km, which correspond to visual geometric albedos of 0.08 and 0.18, respectively. Note that these values are both substantially lower than Charon's albedo, which questions the validity of our assumed phase coefficient. However, we reiterate that the albedo of Hydra is effectively unconstrained at the high end due to the uncertainty in our mass solution. If we reduce the mass of Nix by 1$\sigma$ and assume a density of 2 gm cm$^{-3}$, then its diameter would decrease to 41 km and the albedo would increase to 0.39, so we cannot rule out a Charon-like albedo for either satellite. However, if the densities and compositions are the same, it becomes hard to reconcile large albedo differences given that the evolution of all the satellite surfaces should be similar.

On the other hand, if we assume that Nix and Hydra have albedos similar to Charon we derive diameters of 44 and 52 km, respectively. These diameters imply unreasonably large densities of 13.3 and 4.4 gm cm$^{-3}$, respectively, using the best-fit values of their masses. Clearly, it makes more cosmochemical sense to assume a Charon-like density than to assume a Charon-like albedo for the new satellites. However, the lower limits on mass for both satellites are poorly constrained, so reasonable densities are still allowed by the data, even with such a high albedo. We don't expect the densities to be the same for all objects. After all, the mass and thus self-gravity for Charon is much higher than for Nix and Hydra and should lead to a higher density. By this reasoning one might expect lower densities for the smaller bodies, but a reasonable explanation for such high densities is hard to fathom.

The situation is illustrated in Fig. 1, where the solid curves correspond to the nominal mass solutions, and the dashed curves correspond to the ± 1$\sigma$ mass solutions. The data are consistent with a negligible mass for Hydra and is thus not very constraining. However, we have detected the mass of Nix at the 1$\sigma$ confidence level, and it is large enough to argue that both its density and albedo are less likely to be the same as Charon's. Therefore a simple scenario for a similar composition and internal structure of the three satellites would seem to be unlikely even at this early stage of knowledge about these objects. Also, to first order, we would expect Nix and Hydra to be more similar to one another than to either Pluto or Charon, given that evolutionary processes, such as radiation darkening of their surfaces, are likely to be more similar for such small objects, in comparison to Pluto's surface, where a dynamic interaction with its atmosphere is taking place.

Fig. 1

*Charon / Pluto mass ratio.* The Charon / Pluto mass ratio falls out quite naturally from the four-body orbit solution. Because our best-fit value of 0.1166 is derived from the same data used by B06, it is not surprising that our result is virtually identical to theirs, though with a somewhat larger uncertainty, most likely the result of the increased astrometric uncertainties that we assigned to the observations. Previous authors have struggled to determine this quantity by looking at the barycentric wobble of the system. A chronology of mass ratio determinations is shown in Table 6.

Table 6

*Charon's orbit.* Note that the eccentricity for Charon disagrees with the zero eccentricity published in B06. That erroneous value was the result of two factors. First, the software they used to fit the orbit of Charon had been developed to fit highly eccentric orbits of Kuiper belt binaries such as (58534) 1997 $CQ_{29}$ and (66652) 1999 $RZ_{253}$ (Noll *et al.* 2004a, 2004b), and so it utilized a set of orbital elements that included the date of periapsis, a parameter that becomes meaningless for circular orbits. By itself, the use of that parameter would not have caused the problem. However, they also used the TB97 orbit of Charon as a source of initial values for the fitting routine. The TB97 orbit placed the longitude of periapsis on almost the opposite side of the orbit from where the newer data now place it. To get to the current longitude, the fitting program therefore needed to rotate the line of apsides through almost 180 deg from where it started. In its effort to move periapsis to the opposite side of the orbit, the fitting routine shrank the eccentricity to zero, but once it became zero, the date of periapsis became meaningless and was ignored, preventing the routine from being able to rotate the line of apsides as needed. The code has since been modified to make use of a different set of orbital elements, more suitable for circular and near-circular orbits (Grundy *et al.* 2007). Using that new code, an eccentricity identical to what is tabulated in Table 3 is found, and the argument of periapsis differs by only 0.1 deg.

Unfortunately, the zero eccentricity of Charon's orbit from B06 led Lee & Peale (2006) to suggest that masses near the upper end of the expected range for Nix and Hydra might already be ruled out, due to the eccentricity that would otherwise be induced in Charon's orbit. Our eccentricity of 0.0035 is considerably larger than the amount that would be induced by high-mass satellites (see Fig. 11 of Lee & Peale), so it would appear that the eccentricity of Charon's orbit will not serve to place a more stringent upper limit on the masses of the new satellites than what would be cosmochemically reasonable.

Interestingly, the magnitude of the eccentricity in Charon's orbit is very similar to the value found by TB97 when an early-generation Pluto surface albedo model was applied to the astrometry, in an attempt to correct for the offset between the center of light and the center of mass. However, the new determination of the argument of periapsis for Charon differs from the TB97 result by over 160 deg, as noted above. We considered the possibility that perturbations by Nix and Hydra were responsible for causing the line of apsides to precess at a rate of about 16 deg $yr^{-1}$, which would explain the discrepancy; however, Fig. 2 shows this speculation to be unfounded. The actual rate of precession is more like $0.030 \pm 0.009$ deg $yr^{-1}$ and appears to be directly proportional to the combined mass of Nix and Hydra, so even if we force their masses to their reasonable upper limits, the rate of precession is inadequate to explain the discrepancy in the argument of periapsis. A more likely explanation for the discrepancy is the immaturity of the surface albedo models used to compute the center of light offset from the center of mass for the 1992-1993 data.

Fig. 2

In the Data section, we mentioned three possible reasons for the discrepancy between the semimajor axes for Charon derived from the 1992-1993 and 2002-2003 data. With the four-body orbit solution in hand, we are in a position to address one of those possibilities. A fifty-year integration shows variation in the semimajor axis of Charon's orbit of less than a kilometer, far too little to explain the discrepancy. As for the other two possibilities, the image scales determined for both the Planetary Camera used for the 1992-1993 data and the ACS HRC used for the 2002-2003 data have a strong heritage, and we have no reason to doubt either of them. Once again, we are left with questioning the validity of the center of light offsets from the center of mass used to correct the 1992-1993 data. Now that improved surface albedo models are available, it is possible to repeat the computation of the correction applied to the older data and recompute the orbit, a task we leave for the future. It should be noted that the 2002-2003 data do not suffer from the same problem, because in the process of extracting surface albedo maps from the higher spatial resolution ACS HRC data, a solution for the location of the center of disk (assumed to coincide with the center of mass) was explicitly performed.

*Nix's and Hydra's orbits.* Our best-fit solution shows that the orbits of Nix and Hydra are not quite coplanar with the orbit of Charon. A long integration of the system therefore shows the resulting precession of their poles around the pole of the system's invariable plane. The pole of Nix's orbit is offset from the pole of the invariable plane by $0.15 \pm 0.04$ deg, and the period of precession is slightly less than 5 years. The offset for the pole of Hydra's orbit is slightly larger at $0.19 \pm 0.03$ deg, but the precession period is about three times longer than for Nix. To conserve orbital angular momentum, the orbit plane for Charon must also precess around the pole of the system's invariable plane, though because it has over 1000 times the combined mass of Nix and Hydra, the offset is only 0.0002 deg. We illustrate this behavior in Fig. 3, where the osculating inclination is plotted against time for the three satellites. A similar plot of the osculating longitude of the ascending node is essentially identical, other than the numerical values of the ordinate and the phase, and is therefore not shown. Taken together, the poles of Nix's and Hydra's orbit planes describe circles with radii of $0.15 \pm 0.04$ and $0.19 \pm 0.03$ deg, respectively, around the pole of the invariable plane (effectively the pole of Charon's orbit).

Fig. 3

Figure 4 shows the time evolution of the barycentric osculating eccentricity for Nix and Hydra. The 6.4 day modulation evident in both cases is obviously due to Charon. The slower modulation reflects the orbital period of the satellite in question.

Fig. 4

*Resonance.* B06's two-body orbit solutions indicated that the 4:1 and 6:1 mean-motion resonances were formally excluded for Nix and Hydra, respectively. It is natural to ask whether the four-body orbit solution confirms this result or not. A 50 year integration of our best-fit solution as well as the 25 variant $1\sigma$ solutions did not reveal any instances of resonance among the 59 resonant arguments that we investigated. The average ratios of the osculating orbital periods for Hydra, Nix, and Charon are $6.064 \pm 0.006 : 3.991 \pm 0.007 : 1$. The period ratio for Hydra to Nix is $1.519 \pm 0.014$.

Lee & Peale (2006) found that the resonant argument $\theta = 2\phi_N - 3\phi_H + \varpi$ circulated for both satellites if their masses were small; however, for the large mass case, the argument circulated for Nix but librated around 180 deg for Hydra. We imposed their large mass values (0.18 km$^3$ sec$^{-2}$ for Nix and 0.33 km$^3$ sec$^{-2}$ for Hydra, when expressed in GM units) on our solution and found a $\chi^2$ statistic over $7\sigma$ larger than our best-fit solution. We can therefore say that the observations can rule out that particular extreme case.

*Ephemeris uncertainty.* We integrated our 25 variant solutions from 1980 to 2030 and examined the cloud of ephemeris positions for each satellite on the plane of the sky. The radius of a circle with the same effective area enclosed by the convex hull around each cloud is shown as a function of time in Fig. 5.  The high-frequency structure in each curve is an artifact of the small number of ephemeris positions used to create the cloud.  The smoothed behavior is about what one would expect, with a minimum around the time of the observations, indicated by the horizontal bar, and increasing with time both before and after.  Because the Pluto system was viewed nearly edge-on in the late 1980s, the areas of the skinny ephemeris uncertainty ellipses are smaller than for 15 years into the future, which explains why the uncertainty grows more rapidly in the immediate future than in the immediate past.  The current ephemeris uncertainties are about twice the angular diameters of Nix and Hydra, which would need to be taken into account when predicting stellar occultations by either object.

Fig. 5

Finally, in Fig. 6 we show the difference between satellite ephemerides computed using two-body orbit solutions and our four-body orbit solution.  As expected, departures from unperturbed motion are larger for Nix, which is closer to the primary perturber, Charon.

Fig. 6

## 7. SUMMARY

We have performed a four-body orbit solution for the Pluto system using published observations of the newly discovered satellites Nix and Hydra.  We have established useful 1$\sigma$ upper limits on the masses of Nix and Hydra that place constraints on how low their geometric albedos and how high their densities can be.

The orbits of Charon, Nix, and Hydra are not quite coplanar, with the result that the latter satellites' orbit planes precess around the system's invariable plane with periods of 5 to 15 years.  The orbital eccentricities are nonzero but small for all three satellites when measured in a barycentric reference frame.  The line of apsides for Charon's orbit precesses, but at a rate too slow to account for the difference in the longitudes of periapsis seen in the 1992-1993 and 2002-2003 data sets.  We believe that the offset between the center of light and the center of mass in the 1992-1993 data is responsible for both an incorrect longitude of periapsis and semimajor axis for Charon's orbit.  We see no evidence of any mean motion resonances between the three satellites of Pluto.

## ACKNOWLEDGMENTS

This work was supported by grant HST-AR-10940.02-A.  GTE acknowledges support from the NSF Research Experiences for Undergraduates program.  Our understanding of the four-body problem benefitted from discussions with Bob Jacobson.

REFERENCES

Beletic, J. W., Goody, R. M., & Tholen, D. J. 1989, Icarus, 79, 38
Bevington, P. R. 1969, Data Reduction and Error Analysis for the Physical Sciences (McGraw-Hill)
Buie, M. W., Grundy, W. M., Young, E. F., Young, L. A., & Stern, S. A. 2006, AJ, 132, 290
Buie, M. W., Tholen, D. J., & Wasserman, L. H. 1997, Icarus, 125, 233
CODATA 2006, http://physics.nist.gov/cgi-bin/cuu/Value?bg|search_for=G
Everhart, E. 1985, Dynamics of Comets: Their Origin and Evolution, Proceedings of IAU Colloq. 83, Astrophysics and Space Science Library, 115, 185
Foust, J. A., Elliot, J. L., Olkin, C. B., McDonald, S. W., Dunham, E. W., Stone, R. P. S., McDonald, J. S., & Stone, R. C. 1997, Icarus, 126, 362
Fuentes, C. I., & Holman, M. J. 2006, IAU Electronic Telegram 602
Grundy, W. M., Stansberry, J. A., Noll, K. S., Stephens, D. C., Kern, S. D., Trilling, D. E., Spencer, J. R., Cruikshank, D. P., & Levison, H. F. 2007, Icarus, 191, 286
Gulbis, A. A. S., Elliot, J. L., Person, M. J., Adams, E. R., Babcock, B. A., Emilio, M., Gangestad, J. W., Kern, S. D., Kramer, E. A., Osip, D. J., Pasachoff, J. M., Souza, S. P., & Tuvikene, T. 2006, Nature 439, 48
Lee, M. H., & Peale, S. J. 2006, Icarus, 184, 573
Noll, K. S., Stephens, D. C., Grundy, W. M., & Griffin, I. 2004a, Icarus, 172, 402
Noll, K. S., Stephens, D. C., Grundy, W. M., Osip, D. J., & Griffin, I. 2004b, AJ, 128, 2547
Null, G. W., & Owen, W. M., Jr. 1996, AJ, 111, 1368
Null, G. W., Owen, W. M., Jr., & Synnott, S. P. 1993, AJ, 105, 2319
Olkin, C. B., Wasserman, L. H., & Franz, O. G. 2003, Icarus, 164, 254
Person, M. J., Elliot, J. L., Gulbis, A. A. S., Pasachoff, J. M., Babcock, B. A., Souza, S. P., & Gangestad, J. 2006, AJ, 132, 1575
Press, W. H., Flannery, B. P., Teukolsky, S. A., & Vetterling, W. T. 1986, Numerical Recipes (Cambridge University Press)
Reinsch, K., Burwitz, V., & Festou, M. C. 1994, Icarus, 108, 209
Sicardy, B., Ageorges, N., Marco, O., Roques, F., Mousis, O., Rousselot, P., Hainaut, O., Bellucci, A., Colas, F., Gendron, E., Lellouch, E., Renner, S., & Widemann, T. 2006a, IAU Electronic Telegram 610
Sicardy, B., Bellucci, A., Gendron, E., Lacombe, F., Lacour, S., Lecacheux, J., Lellouch, E., Renner, S., Pau, S., Roques, F., Widemann, T., Colas, F., Vachier, F., Vieira Martins, R., Ageorges, N., Hainaut, O., Marco, O., Beisker, W., Hummel, E., Feinstein, C., Levato, H., Maury, A., Frappa, E., Gaillard, B., Lavayssière, M., Di Sora, M., Mallia, F., Masi, G., Behrend, R., Carrier, F., Mousis, O., Rousselot, P., Alvarez-Candal, A., Lazzaro, D., Veiga, C., Andrei, A. H., Assafin, M., da Silva Neto, D. N., Jacques, C., Pimentel, E., Weaver, D., Lecampion, J.-F., Doncel, F., Momiyama, T., & Tancredi, G. 2006b, Nature, 439, 52
Stern, S. A., Mutchler, M. J., Weaver, H. A., & Steffl, A. J. 2007, AJ, submitted
Tholen, D. J., & Buie, M. W. 1988, AJ, 96, 1977
Tholen, D. J., & Buie, M. W. 1990, BAAS, 22, 1129
Tholen, D. J., & Buie, M. W. 1997, Icarus, 125, 245

Ward, William R., & Canup, R. M. 2006, Science, 313, 1107

Weaver, H. A., Stern, S. A., Mutchler, M. J., Steffl, A. J., Buie, M. W., Merline, W. J., Spencer, J. R., Young, E. F., & Young, L. A. 2005, IAU Circular 8625

Weaver, H. A., Stern, S. A., Mutchler, M. J., Steffl, A. J., Buie, M. W., Merline, W. J., Spencer, J. R., Young, E. F., & Young, L. A. 2006, Nature, 439, 943

Young, E. F., & Binzel, R. P. 1994, Icarus, 108, 219

Young, L. A., Olkin, C. B., Elliot, J. L., Tholen, D. J., & Buie, M. W. 1994, Icarus, 108, 186

Young, L. A., Olkin, C. B., Young, E. F., French, R. G., Shoemaker, K., Ruhland, C.. Gregory, B., & Galvez, R. 2005, IAU Circular 8570

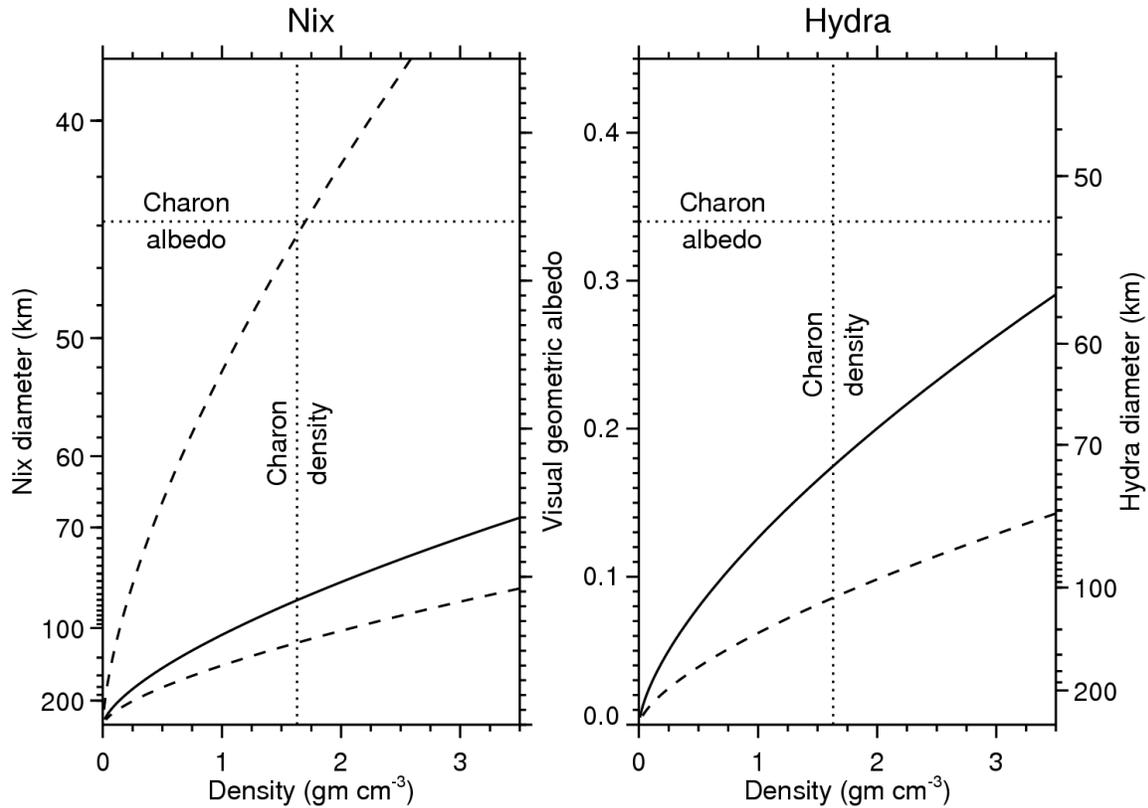

Fig. 1  Relationship between diameter, albedo, and density for Nix (left) and Hydra (right).  For the nominal masses in Table 2, the solid curves indicate how higher densities correspond to higher albedos and smaller diameters.  The dashed curves show the same for the nominal masses plus 1σ (and for Nix only, minus 1σ).  The albedo and density of Charon are indicated by horizontal and vertical dotted lines.  Our nominal masses are not compatible with Nix and Hydra having the same density and albedo as Charon.  They would require Nix and Hydra albedos to be lower or their densities higher. Our 1σ mass uncertainties do still allow Charon-like properties, although only barely in the case of Nix.  If Nix has a low density, as recently reported for some trans-Neptunian objects, then its albedo is likely lower than that of Charon.

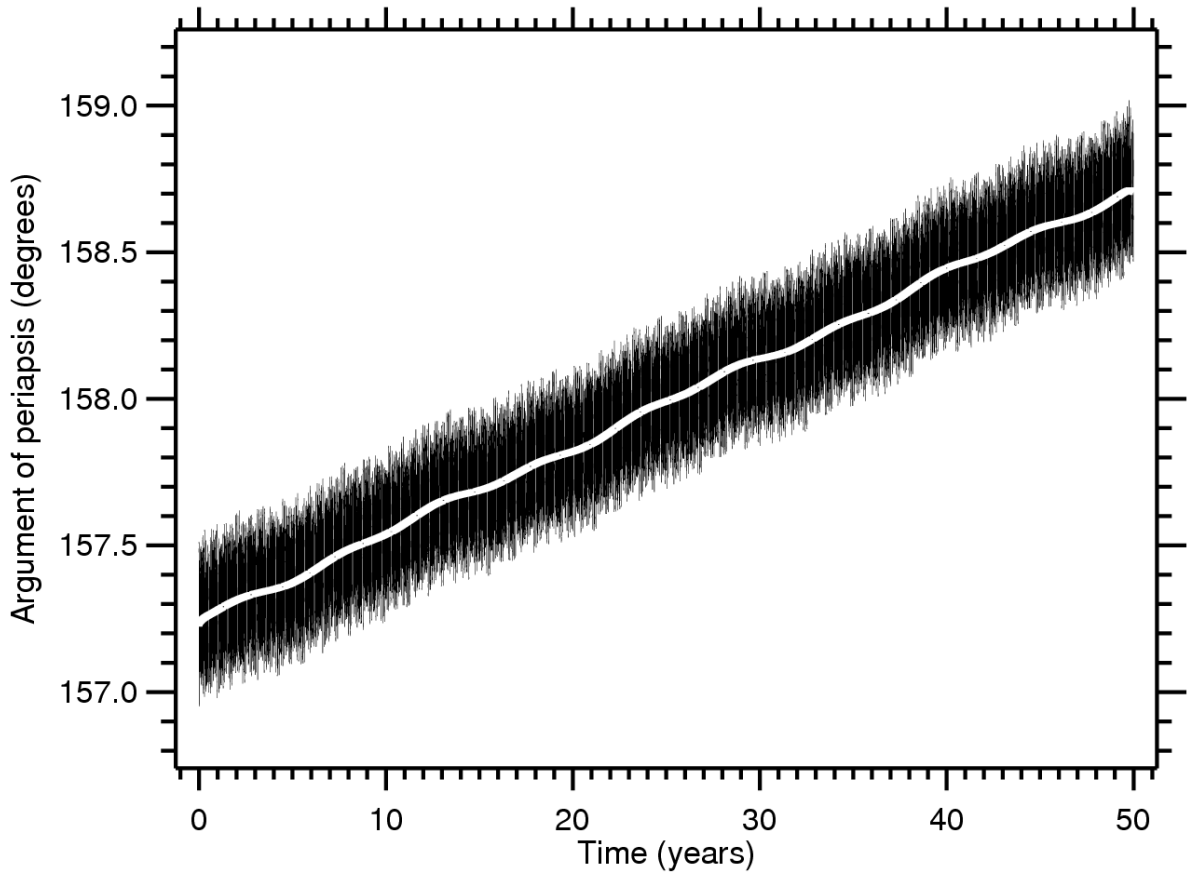

Fig. 2  Argument of periapsis for Charon versus time. This plot shows the osculating argument of periapsis of Charon over a 50-year period, starting on 1980 January 01 UT. The superimposed white line is the time-averaged behavior with a smoothing width of 0.5 yr. The rate of precession is 0.030 deg yr$^{-1}$ and corresponds to a period of 12000 years. The fastest precession rate among the 1$\sigma$ orbits is 0.048 deg yr$^{-1}$, and the slowest is 0.011 deg yr$^{-1}$.

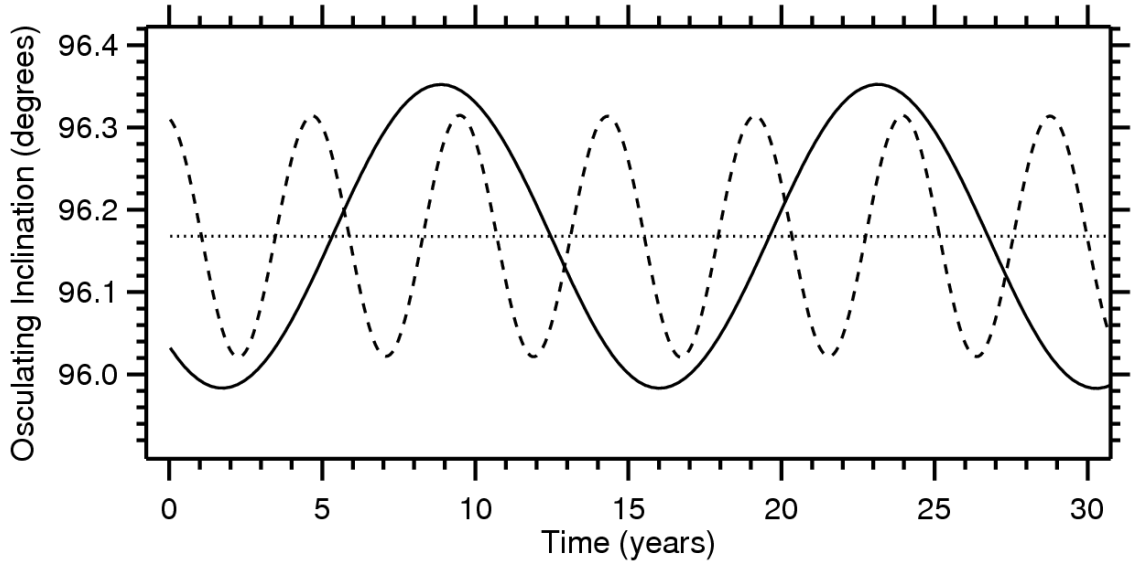

Fig. 3  Variation of the osculating inclination with time.  This figure shows the slow variation of inclination with time (arbitrary zero point) for the three satellites.  The variation is shown by the solid curve for Hydra and by the dashed curve for Nix.  Both curves oscillate around the inclination of Charon (dotted curve).  The osculating inclinations vary by ± 0.0002 deg for Charon, ± 0.15 deg for Nix, and ± 0.19 deg for Hydra.

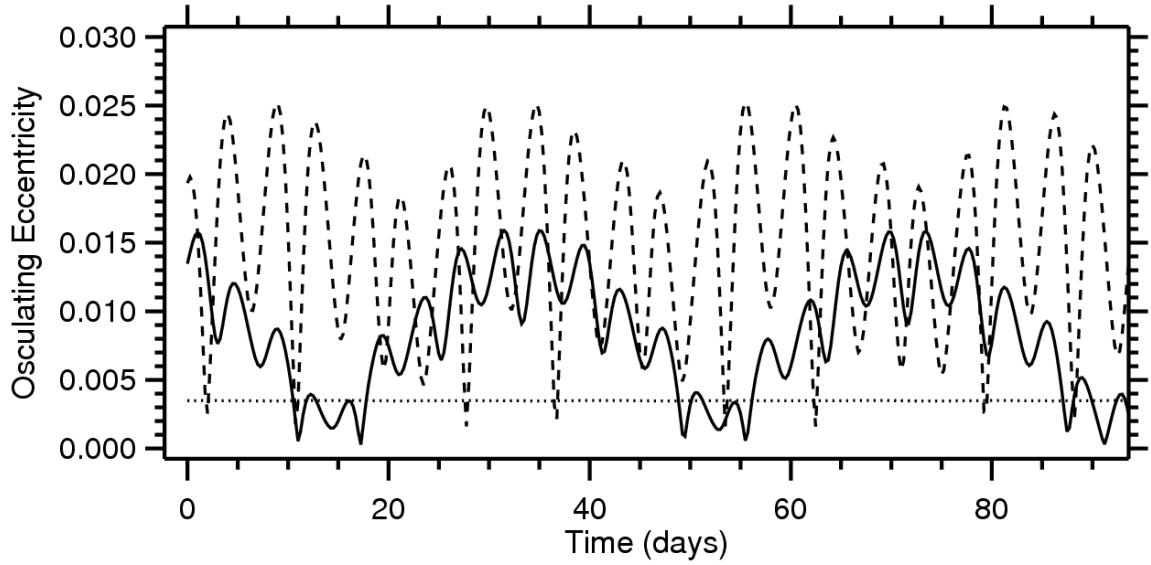

Fig. 4  Osculating eccentricity as a function of time.  This plot shows the rapidly varying nature of the osculating eccentricity of Nix and Hydra with time (arbitrary zero point).  Hydra is plotted with the solid curve, the dashed curve shows the behavior of Nix, and the dotted curve is for Charon.  The value for Charon's eccentricity varies by only $2 \times 10^{-5}$.  The variation seen for Nix is [0, 0.0272] and for Hydra is [0, 0.0179].  The fast variation in the eccentricities is tied to the orbital period of Charon.  The slower modulation is tied to the orbital period of the respective satellite.

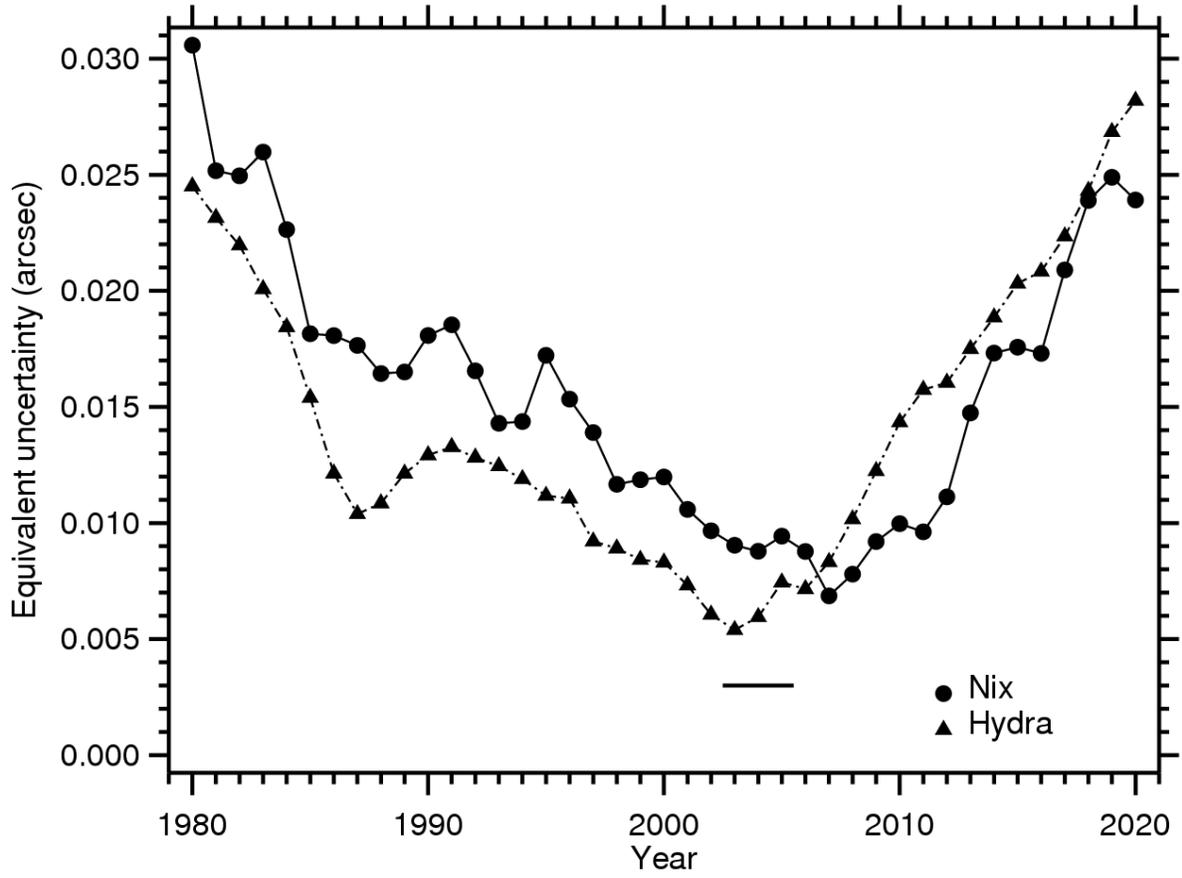

Fig. 5 Sky-plane positional uncertainty versus time based on the current dataset. This figure shows an estimate of the total sky-plane uncertainty for Nix (filled circles) and Hydra (triangles). The solid bar below the curves shows the time-span of the astrometry that constrains the orbit solution.

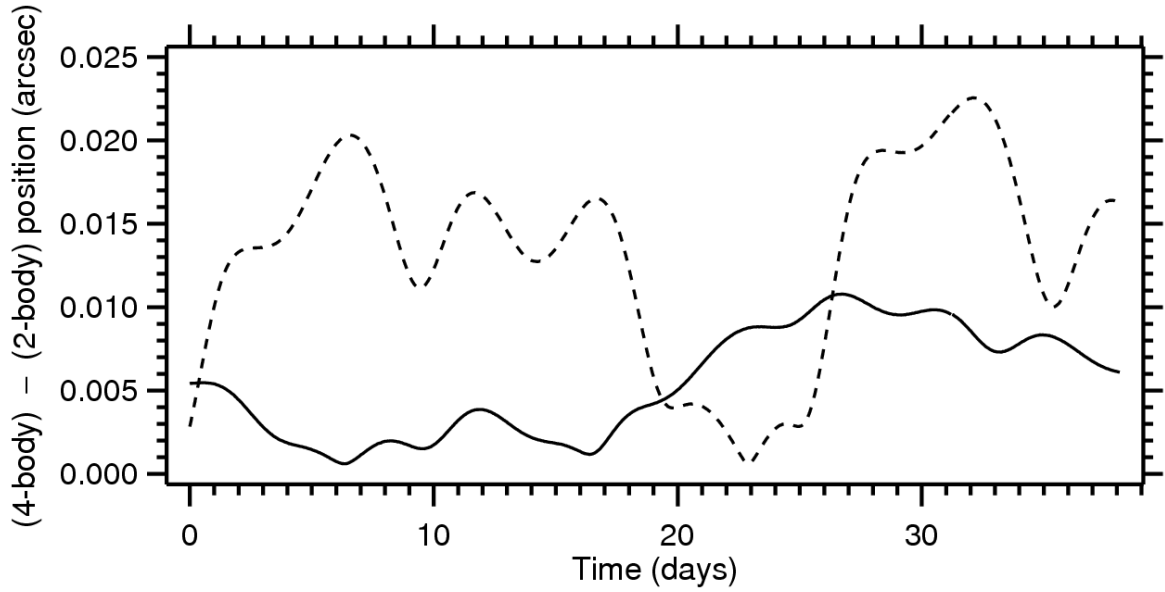

Fig. 6  Offset between four-body orbit fit and two-body orbit fits in 2003.  This figure shows the difference between the two orbit fits over the timespan of one Hydra orbit period (arbitrary time zero point).  The solid line is for Hydra and the dashed line is for Nix. The departure from unperturbed motion is clearly larger for the closer satellite, Nix.

TABLE 1
State vectors, epoch JD 2452600.5, mean equator and equinox of J2000

|       | x (km)    | y (km)    | z (km)   | x′ (km/day) | y′ (km/day) | z′ (km/day) |
|-------|-----------|-----------|----------|-------------|-------------|-------------|
| Charon | 3.8<br>−12614.0 | 4.6<br>−10150.0 | 2.9<br>11061.1 | 2.8<br>6842.2 | 3.5<br>8715.1 | 1.5<br>15699.2 |
| Nix   | 180.<br>8450. | 210.<br>1020. | 100.<br>−46480. | 28.<br>−7930. | 39.<br>−7533. | 54.<br>−542. |
| Hydra | 170.<br>2120. | 180.<br>11830. | 68.<br>64674. | 15.<br>8479. | 19.<br>7900. | 22.<br>−132. |

TABLE 2
GM values, masses, densities, diameters, and albedos for the Pluto system

|  | GM (km$^3$ sec$^{-2}$) | mass (kg) | density (gm cm$^{-3}$) | diameter (km) | albedo (V) |
|---|---|---|---|---|---|
| Pluto | 870.3 ± 3.7 | $1.304 \times 10^{22}$ | 2.06 | [2294] | 0.61 |
| Charon | 101.4 ± 2.8 | $1.520 \times 10^{21}$ | 1.63 | 1212 | 0.34 |
| Nix | 0.039 ± 0.034 | $5.8 \times 10^{17}$ | [1.63] | 88 | 0.08 |
| Hydra | 0.021 ± 0.042 | $3.2 \times 10^{17}$ | [1.63] | 72 | 0.18 |
| total | 971.78 ± 0.20 | $1.456 \times 10^{22}$ | 2.01 | — | — |

TABLE 3
Keplerian orbital elements, epoch JD 2452600.5, mean equator and equinox of J2000

|  | a (km) | e | i (deg) | Ω (deg) | ω (deg) | L (deg) | P (days) |
|---|---|---|---|---|---|---|---|
| Charon [a] | 19570.3 | 0.0035 | 96.168 | 223.054 | 157.9 | 257.960 | 6.38720 |
| Nix [b] | 49240. | 0.0119 | 96.190 | 223.202 | 244.3 | 122.7 | 25.49 |
| Hydra [b] | 65210. | 0.0078 | 96.362 | 223.077 | 45.4 | 322.4 | 38.85 |

[a] Plutocentric
[b] barycentric

TABLE 4
Mean orbital elements, mean equator and equinox of J2000

| | a (km) | e | i (deg) | Ω (deg) | ω (deg) | P (days) |
|---|---|---|---|---|---|---|
| Charon[a] | 19570.45 ± 44 | 0.003484 ± 36 | 96.1680 ± 28 | 223.0539 ± 32 | 157.92 ± 32 | 6.387206 ± 7 |
| Nix[b] | 49242. ± 12 | 0.01504 ± 14 | Charon ± 0.15 | | — | 25.492 ± 9 |
| Hydra[b] | 65082. ± 9 | 0.00870 ± 27 | Charon ± 0.19 | | — | 38.734 ± 8 |

[a] Plutocentric
[b] barycentric

TABLE 5
Four-body orbit solution residuals for Nix and Hydra (arcsec)

|  | Nix | | Hydra | |
| --- | --- | --- | --- | --- |
| UTC | Δ R.A. cos δ | Δ Dec. | Δ R.A. cos δ | Δ Dec. |
| 2002 Jun 11 | +0.0037 | −0.0074 | +0.0089 | −0.0048 |
| 2002 Jun 14 | +0.0111 | +0.0137 | +0.0017 | +0.0072 |
| 2002 Jun 18 | −0.0035 | −0.0153 | (+0.1684) | (+0.0171) |
| 2002 Jul 02 | −0.0180 | +0.0054 | −0.0128 | −0.0160 |
| 2002 Jul 17 | −0.0203 | +0.0135 | +0.0009 | +0.0027 |
| 2002 Oct 03 | −0.0148 | −0.0457 | −0.0202 | −0.0121 |
| 2003 Feb 18 | −0.0211 | −0.0131 | +0.0070 | −0.0034 |
| 2003 Apr 20 | −0.0073 | +0.0022 | −0.0031 | +0.0012 |
| 2003 May 13 | +0.0204 | −0.0051 | +0.0127 | −0.0058 |
| 2003 May 28 | +0.0054 | +0.0050 | +0.0071 | +0.0145 |
| 2003 May 30 | +0.0147 | +0.0261 | −0.0025 | +0.0081 |
| 2003 Jun 08 | +0.0231 | +0.0004 | +0.0042 | +0.0120 |
|  |  |  |  |  |
| 2005 May 14 | +0.0062 | +0.0241 | −0.0327 | +0.0227 |
| 2005 May 17 | +0.0029 | +0.0205 | +0.0126 | +0.0053 |
|  |  |  |  |  |
| 2006 Feb 15 | −0.0049 | −0.0036 | −0.0006 | +0.0023 |
| 2006 Mar 02 | +0.0027 | +0.0135 | +0.0062 | −0.0112 |
|  |  |  |  |  |
| 2006 Apr 10 | — | — | (+0.0330) | (−0.0132) |
| 2006 Jun 27 | (+0.1519) | (+0.1158) | (+0.1571) | (+0.0656) |
| 2006 Jun 28 | (+0.1140) | (+0.0821) | (+0.0976) | (+0.0895) |

TABLE 6

Chronology of Charon / Pluto Mass Ratio Determinations

| mass ratio | Reference |
|---|---|
| $0.0837 \pm 0.0147$ | Null *et al.* (1993) |
| $0.1566 \pm 0.0035$ | Young *et al.* (1994) |
| $0.124 \pm 0.008$ | Null & Owen (1996) |
| $0.110 \pm 0.060$ | Tholen & Buie (1997) |
| $0.117 \pm 0.006$ | Foust *et al.* (1997) |
| $0.122 \pm 0.008$ | Olkin *et al.* (2003) |
| $0.1165 \pm 0.0055$ | Buie *et al.* (2006) |
| $0.1166 \pm 0.0069$ | this work |